\documentclass[pra,twocolumn,showpacs,amsmath,amssymb]{revtex4}
\usepackage{color}
\usepackage{graphicx}
\usepackage{amsmath}
\graphicspath{{FIGURES}}

\begin{document}

\title{Quantum Monte Carlo Study of the Rabi-Hubbard Model}
\author{T. Flottat$^{1}$, F. H\'ebert$^1$,
  V.G. Rousseau$^2$, and G.G. Batrouni$^{1,3,4,5}$}
\affiliation{$^1$UCA, CNRS, INLN; 1361 route des Lucioles, 06560 Valbonne, France}
\affiliation{$^2$ Physics Department, Loyola University New Orleans, 6363 Saint Charles Ave., LA 70118, USA}
\affiliation{$^3$Institut Universitaire de France, 103 bd Saint-Michel, 75005 Paris, France,}
\affiliation{$^4$MajuLab, CNRS-UNS-NUS-NTU International Joint Research Unit UMI 3654, Singapore}
\affiliation{$^5$Centre for Quantum Technologies, National University of Singapore; 2 Science Drive 3 Singapore 117542}

\begin{abstract}We study, using quantum Monte Carlo (QMC) simulations, the
  ground state properties of a one dimensional Rabi-Hubbard model.
  The model consists of a lattice of Rabi systems coupled by a photon
  hopping term between near neighbor sites. For large enough coupling
  between photons and atoms, the phase diagram generally consists of
  only two phases: a coherent phase and a compressible incoherent one
  separated by a quantum phase transition (QPT). We show that, as one
  goes deeper in the coherent phase, the system becomes unstable
  exhibiting a divergence of the number of photons.  The Mott phases
  which are present in the Jaynes-Cummings-Hubbard model are not
  observed in these cases due to the presence of non-negligible
  counter-rotating terms.  We show that these two models become
  equivalent only when the detuning is negative and large enough, or
  if the counter-rotating terms are small enough.
\end{abstract}


\maketitle

\section{Introduction}

In recent years, the possibility to build \cite{schoelkopf08,blais04}
elementary cavity quantum electrodynamics systems, whether formed by
an atom in a cavity \cite{haroche06} or by a Josephson junction
coupled to microwave photons on a chip \cite{wallraff04}, opened the
perspective to construct new many body systems from such elementary
building blocks.

The Jaynes-Cummings-Hubbard (JCH) model \cite{jaynes63} is the
simplest model describing such an array of coupled cavities. Each
cavity consists of a two-level system (describing the atom or the
junction) coupled to a unique mode of the cavity. The cavities are
then coupled by tunneling of photons from one cavity to the next. In
this many-cavity system, it was predicted that a phenomenon similar to
a photon blockade \cite{birnbaum05} would take place, leading to a
phase transition transition between a coherent state, where all the
cavities are in phase, and an incoherent state where the number of
excitations per cavity is quantized, reminiscent of the
Mott-superfluid transition observed in the bosonic Hubbard model
\cite{greentree06,hartman06,rossini07,koch09,hartmann08,blatter09,zhao08}.
Non equilibrium properties of such systems have also been extensively
discussed \cite{tomadin10,carusotto13}.

In the JCH model description of such a system, the total number of
excitons, $N$, the sum of the number of photons $N_l$ and of excited
atoms (also referred to as spins) $N_s$, is a conserved quantity.
However, a complete description of the interaction between the
two-level system and the photon field should include, as described by
the Rabi Hamiltonian \cite{rabi36}, so-called counter-rotating (CR)
terms that do not conserve the number of excitations.  Neglecting
these terms, which is generally called the rotating wave approximation
(RWA), is generally valid due to the very small values of the coupling
$g$ between the field and the material device. Compared to the energy
of a photon $\omega_l$, $g$ generally ranges from $g/\omega_l \simeq
10^{-7}$ in cavity quantum electrodynamics (QED) experiments up to
$g/\omega_l \simeq 10^{-3}$ in circuit QED \cite{blais04}.

However, it was recently remarked that, for larger values of $g$,
these CR terms may modify deeply the physics of these systems
\cite{zheng11,schiro12,schiro13}.  In \cite{schiro12}, Schiro {\it et
  al.} introduced the Rabi-Hubbard (RH) model, which is equivalent to
the JCH model with the addition of counter-rotating terms, both in the
interaction and the hopping terms. This model has different symmetry
properties from the JCH model and consequently different transitions.
The studies performed in \cite{zheng11,schiro12,schiro13,kumar13} show
that the RH model does not exhibit Mott insulating behavior, but
rather an incoherent/coherent transition which resembles the
superradiant transition of the Dicke model
\cite{hepp73,rotondo15}. The out of equilibrium behavior
  of the RH model has been recently studied in
  Ref.\cite{schiro16}. The related Dicke model and the nature of its
  transition are also the subject of intense research
  \cite{rotondo15,zhu16,nataf10,zou14}.  Recent experiments show the
breakdown of the RWA approximation by reaching an ultra strong
coupling regime in circuit QED \cite{niemczyk10,forndiaz16} where
$g/\omega_l \simeq 10^{-1}$.

In order to study the differences between the RH and JCH
  descriptions of these systems, we will use exact quantum Monte Carlo
  (QMC) simulations.  Whereas the JCH model has been studied with DMRG
  \cite{rossini07} and QMC \cite{zhao08} the RH model has mostly been
  studied with mean-field approximations
  \cite{zheng11,schiro12,schiro13}.  Kumar {\it et al.} \cite{kumar13}
  mapped the RH model onto the quantum Ising model in the strong
  coupling limit ($g/\omega_l > 1$) and compared their results with
  DMRG simulations. They found that the QPT of the RH model is well
  described by the dynamics of the quantum Ising model. However, their
  calculation was done with the number of photons per cavity
  restricted to be at most equal to $3$. As we will see below, the
  number of photons per cavity can far exceed this value in the
  coherent phase. In addition, the DMRG calculation of
  Ref. \cite{kumar13} was done without the CR terms in the
  hopping. Below, we will study the effect of these terms.

The JCH and RH models we will study in this paper are but
  two of a very wide variety of related models describing a large
  spectrum of interesting physical situations of current interest. For
  example, Ref. \cite{kurcz14} introduces a model of spins coupled to
  photons which are exchanged at all distances and where the photon
  band is {\it flat}, which is a key difference with the model we
  address here. Another interesting class of models is that where the
  band is not flat and where, in addition, spin-photon coupling is
  mode dependent and chosen to lead to an Ohmic spectral density at
  low enough frequencies \cite{diaz16}. These models can also be
  studied with the QMC algorithm we use in this paper.

In Section II, we review the model, the technique used to simulate it
and the physical quantities of interest for this study. In Section
III, we present the phase diagram of the model and compare it with
previously known results, focusing on the case where the detuning is
zero.  In Section IV, we discuss in more detail the importance of the
counter-rotating terms and their effects.

\section{Models and techniques}

\begin{figure}
\includegraphics[width=8.8cm]{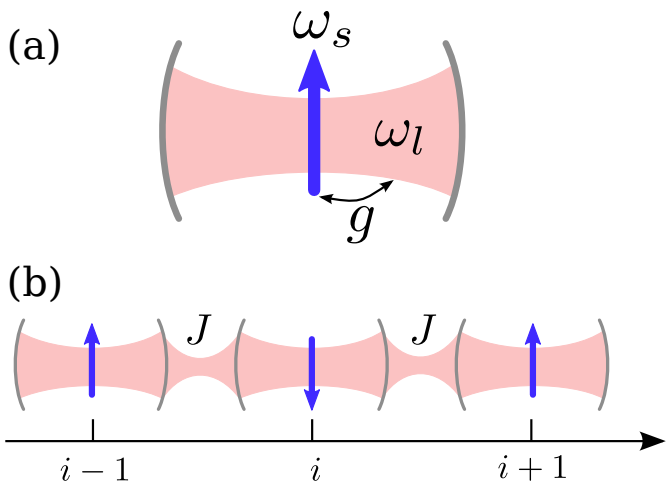}
\caption{(Color online) The Rabi model (a) describes the interaction
of a two levels quantum system (here represented by a spin) with a mode of a photon field
(here in pink). The Jaynes-Cummings or Rabi-Hubbard models describe the coupling
of such Rabi cavities by tunnel effect of the photons between a cavity and the next.}
\label{schema}
\end{figure}

We study the Rabi-Hubbard model on a one-dimensional periodic chain
with $L$ sites. Each site of the chain, labelled with index $i$,
represents a cavity (Fig. \ref{schema}). In each cavity there is a
two-level atom represented as a spin-1/2 and we consider only one
photon mode per cavity.  The atom is coupled to the mode with the
Rabi-Hamiltonian \cite{rabi36}.  The sites are coupled to each other
by a photon hopping term connecting near neighbor cavities via the
tunnel effect.  The Rabi-Hubbard (RH) Hamiltonian governing this
system is \cite{schiro12}
\begin{eqnarray}
  H =&& -J \sum_i(a^{\phantom\dagger}_i + a^\dagger_i)
  (a^{\phantom\dagger}_{i+1} + a^\dagger_{i+1}) \nonumber\\
  &&+ \sum_i \left(\omega_s \sigma^+_i \sigma^-_i + \omega_l
    a^\dagger_i a^{\phantom\dagger}_i \right)
  \nonumber\\
  &&+ g \sum_i (\sigma^-_i + \sigma^+_i)(a^{\phantom\dagger}_i +
  a^\dagger_i) \label{Hbasic}
\end{eqnarray}
The $a_i (a^\dagger_i)$ describes the destruction (creation) of
photons in the $i$-th cavity. The $\sigma_i$ operators describe the
two-level atoms.  The first term in Eq.(\ref{Hbasic}) describes the
hopping of photons between cavities; the second term, which is
diagonal, describes the energies of the photons and the excited atomic
states. The third term describes the exchange between the atom and the
photons field. We will call $\delta = \omega_a - \omega_l$ the detuning between
the atomic and light frequencies

Written in this form, the Hamiltonian includes counter-rotating terms
(CR): terms which create two photons ($a^\dagger_i$ $a^\dagger_{i+1}$)
in the first line of Eq.{(\ref{Hbasic})} or which simultaneously
excite a spin and create a photon ($\sigma^+_i a^\dagger_i$) in the
third, and their Hermitian conjugates.  This last term is often
neglected in the treatment of the Rabi model as it couples states with
different diagonal energies. The non-rotating terms in the hopping
part of the Hamiltonian are probably not physical but were studied in
\cite{schiro12} and allow us to study the effects of two different non
rotating terms.  Ignoring these terms in Eq.(\ref{Hbasic}), what is
generally called the rotating wave approximation (RWA), transforms the
model into the Jaynes-Cummings-Hubbard (JCH) Hamiltonian
\cite{jaynes63}. There is an important qualitative difference between
these models. The JCH model conserves the number of excitations $N =
N_s + N_l$ and has U(1) symmetry whereas the RH model does not have
such a conservation law and has a simpler $Z_2$ symmetry.

To study the differences between the JCH and RH models, we perform
exact numerical simulations using the Stochastic Green Function (SGF)
quantum Monte Carlo algorithm \cite{rousseau08a,rousseau08b}. This
numerical technique is able to tackle the non-conservation of the
number of excitations and the large number of photons present in the
RH model. This allowed us the study its phase diagram and simulate
sizes up to $L=30$ sites and inverse temperatures $\beta \omega_l
\simeq 20$. We chose to study a one dimensional lattice where results
can easily be compared with DMRG studies of the JCH \cite{rossini07}
and RH \cite{kumar13} models and because of the limited number of
sites we can study. To study in more details the effects of the CR
terms, we generalize slightly the forms of the parameters $g$ and $J$.
The hopping term will then be replaced with
\begin{equation}
  -\sum_i J_r (a_i a^\dagger_{i+1} + {\rm h.c.}) + J_{cr} (a_{i} a_{i+1}
  +{\rm h.c.}), 
\label{hoppingdetail}
\end{equation}
and the Rabi term with 
\begin{equation}
  \sum_i g_r  ( \sigma^+_i a_i + \sigma^-_i a^\dagger_i) + g_{cr}
  (\sigma^+_i a^\dagger_i + \sigma^-_i a_i ),
\end{equation}
where we introduce separate parameters for the rotating and
counter-rotating terms ($g_r, J_r$ and $g_{cr},J_{cr}$,
respectively).

We measure the number of photons $N_l$ or the total number of
excitations $N$ and the corresponding densities $n_l=N_l/L$ and $n =
N/L$; we also measure various Green functions
\begin{equation}G_{\alpha\beta}(R) =
\frac{1}{2L}\sum_i\langle \alpha_i \beta_{i+R} +{\rm
  h.c.}\rangle \label{gfunc} 
\end{equation} where $\alpha$ and
$\beta$ are creation or destruction operators for the photons
($a^\dagger$ or $a$) or the spins ($\sigma^+$ or $\sigma^-$). 
This
  allows us to study the phase coherence of photons, spins and coherence
  between photons and spins. For example, we will measure the photon
  equal-time Green function 
\begin{equation}
G_{a^\dagger a}(R) = \frac{1}{2L} \sum_i \langle a^\dagger_i a_{i+R}
+a^\dagger_{i+R} a_i\rangle 
\end{equation}
This yields the photon condensate fraction, $C_l$, which is the order
parameter for the phase coherence of photons:
\begin{equation}
C_l = \frac{\sum_R G_{a^\dagger a}(R)}{N_l} \label{condfraction}.
\end{equation}
Combining different spin-spin correlation functions such as
\begin{equation}
G_{\sigma^-\sigma^+}(R) = \frac{1}{2L} \sum_i \langle \sigma^-_i
\sigma^+_{i+R} +\sigma^-_{i+R} \sigma^+_i\rangle, 
\end{equation}
we obtain the spin-spin correlation
\begin{equation}
C_s = \frac{1}{2L}\sum_i \langle
\sigma^x_i\sigma^x_{i+L/2}\rangle \label{sscorr} 
\end{equation}
at the largest distance $L/2$ in the system which exposes the coupling
of the atoms via the photon field.  

\section{Phases and phase diagram at $\delta=0$}

We start with the Rabi-Hubbard model at $\delta=0$ and with: $g_r =
g_{cr} = g$ and $J_{r} = J_{cr}= J$.  We take $\omega_l = \omega_s =
1$ to fix the energy scale and we study the phase diagram as
$J/\omega_l$ is varied for a fixed value of $g$.

\begin{figure}
    \includegraphics[width=8.8cm]{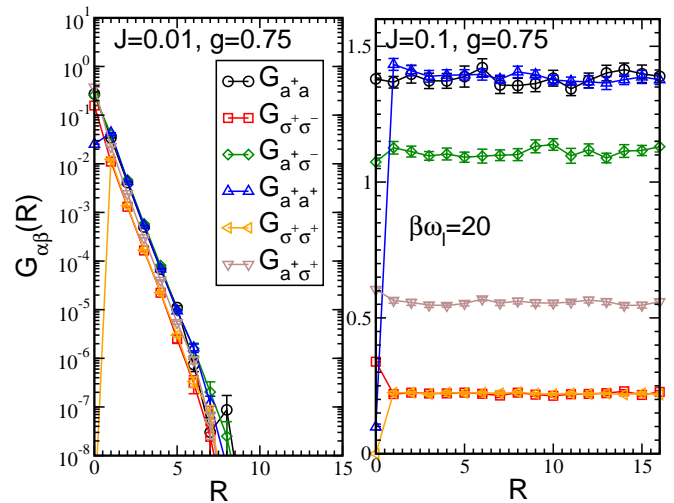}
        \caption{(Color online) Green functions (Eq.(\ref{gfunc})) as a function of distance
in the incoherent (left) and coherent (right) phases. In the incoherent phase, all
the Green functions decay exponentially. In the coherent phase, all the
Green functions show a plateau behavior, reaching a constant value at long
distance. This is due to the discrete symmetry of the Rabi-Hubbard model.
\label{green}}
\end{figure}

We observed, as predicted in \cite{schiro12}, that the
  system undergoes an evolution from an incoherent phase, at low $J$,
  to a coherent one, at large $J$. The incoherent phase is
  characterized by exponential decay of all the Green functions with
  distance indicating the absence of long range and quasi-long range
  order in the system (Fig. \ref{green}, left). This means that the
  photons are not coherent at long distance and that the spins are
  also not ordered.  The density of photons, equal to $G_{a^\dagger
    a}(0)$, is also small in this phase, of the order of
  $3\cdot10^{-1}$ in the case presented here.

  At large $J$ we observe, on the contrary, a phase where all the
  Green functions remain non-zero at large distances
  (Fig. \ref{green}, right).  The photons form a coherent field across
  the system and the spins adopt a ferromagnetic order along the
  $x$-axis. The spins and photons are correlated with each other at
  long distances, as is shown by the behavior of mixed Green functions
  such as $G_{a^\dagger \sigma^-}(R)$.  Finally, the density of
  photons becomes larger (around 1.4 in the present case) in this
  coherent phase. 

\begin{figure}
  \includegraphics[width=8.8cm]{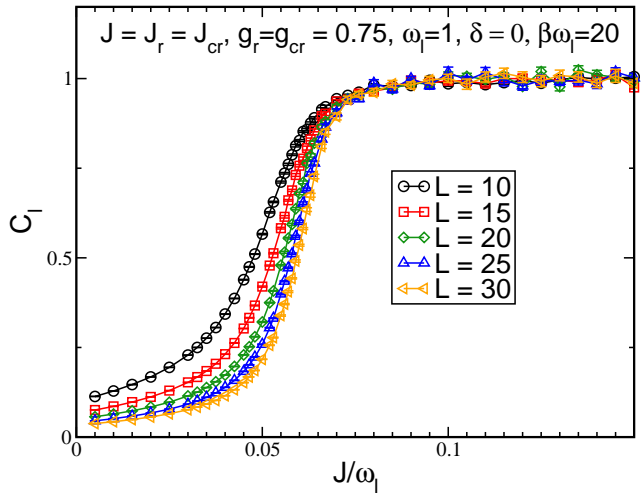}
        \caption{(Color online) Condensate fraction $C_l$
          (Eq.(\ref{condfraction})) of the photons as a function of
          $J/\omega_l$ for detuning $\delta=0$. The system undergoes a
          phase transition from an incoherent field to a coherent
          field as the coupling $J$. 
\label{cond_frac}}
\end{figure}

To study the transition between the incoherent and
  coherent regimes for the photons, we will look at the evolution of
  the condensate fraction $C_l$.  The value of $C_l$
  (Fig. \ref{cond_frac}) is not zero for small $J$ while we observe
  that all the Green functions (see Fig. \ref{green}, left) decrease
  exponentially in this phase. We will show that this non-zero $C_l$
  is only a finite size effect. $C_l$ increases as the system becomes
  coherent at larger $J$.

A striking feature is that $C_l$ almost reaches its saturation value
of unity for large $J$ indicating that all the photons are
condensed. This is atypical in one dimension, where one usually
encounters quasi-condensation of the photons. In the present case, the
condensation is due to the discrete $Z_2$ symmetry of the RH
model. This is also visible in the behavior of the Green functions in
the coherent phase (Fig. \ref{green}, right) where all the Green
functions reach plateaux with constant values at large distances,
instead of the power law decay generally observed in one dimension.
Another important feature is that we observe only the two phases,
incoherent and coherent photon phases, with no sign of any Mott
insulator where $C_l$ would be zero for commensurate densities. This
result confirms the mean field analysis of \cite{schiro12}.

\begin{figure}
	\includegraphics[width=8.8cm]{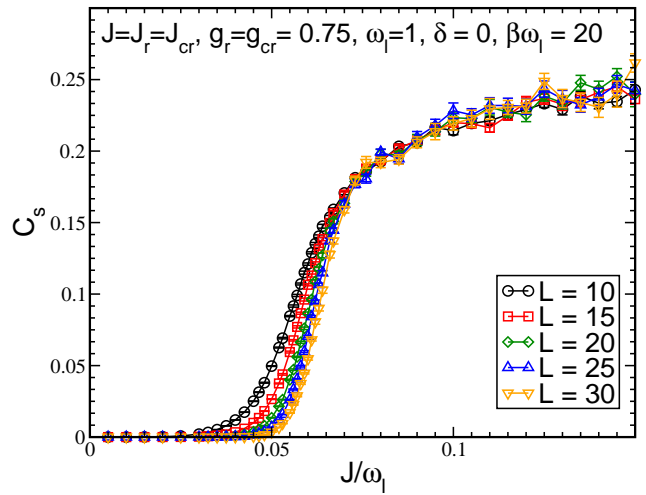}
        \caption{(Color online) Spin-spin correlations $C_s$
          (Eq.(\ref{sscorr})) at the maximum 
          distance $R=L/2$ as a function of $J/\omega_l$ for detuning
          $\delta=0$ and for different sizes.  As the photons become
          coherent (see Fig. \ref{cond_frac}), the system becomes
          ``ferromagnetic".
          \label{ss}}
\end{figure}

The spin-spin correlations $C_s$ follow the behavior of the condensate
fraction (Fig. \ref{ss}): as the photons become coherent, the spins
couple to the photon field and become ferromagnetic (or ferroelectric
in terms of polarization of the atoms). Once again, we find true long
range order for the spin-spin correlations (Fig. \ref{green}) which
tend to their saturated value of $1/4$ when $J/\omega_l$ becomes
large. We observe that both $C_l$ and $C_s$ become large
  simultaneously, which suggests that there is no intermediate phase
  where one species is coherent while the other is not.

  The transition is similar to the Dicke transition \cite{rotondo15}.
  The main difference between the RH and Dicke models is that only one
  photonic mode is present in the latter, while there are several in
  the former.  However (Fig. \ref{green}, right) $G_{a^\dagger a}(R)$
  reaches a plateau at long distance.  There is then a Bose
  condensation of the photons with a macroscopic occupation of the
  $k=0$ mode in Fourier space. Indeed, the number of photons $N(k)$
  occupying a Fourier mode of wave vector $k$ is simply the Fourier
  transform of $G_{a^\dagger a}(R)$.  Then, despite the fact that
  there are many photon modes, the system selects the lowest energy
  one when it becomes coherent and the resulting coherent phase is
  very similar to that in the Dicke system. Similarities between the
  JCH and Dicke model were discussed in \cite{schmidt13}.

\begin{figure}
    \includegraphics[width=8.8cm]{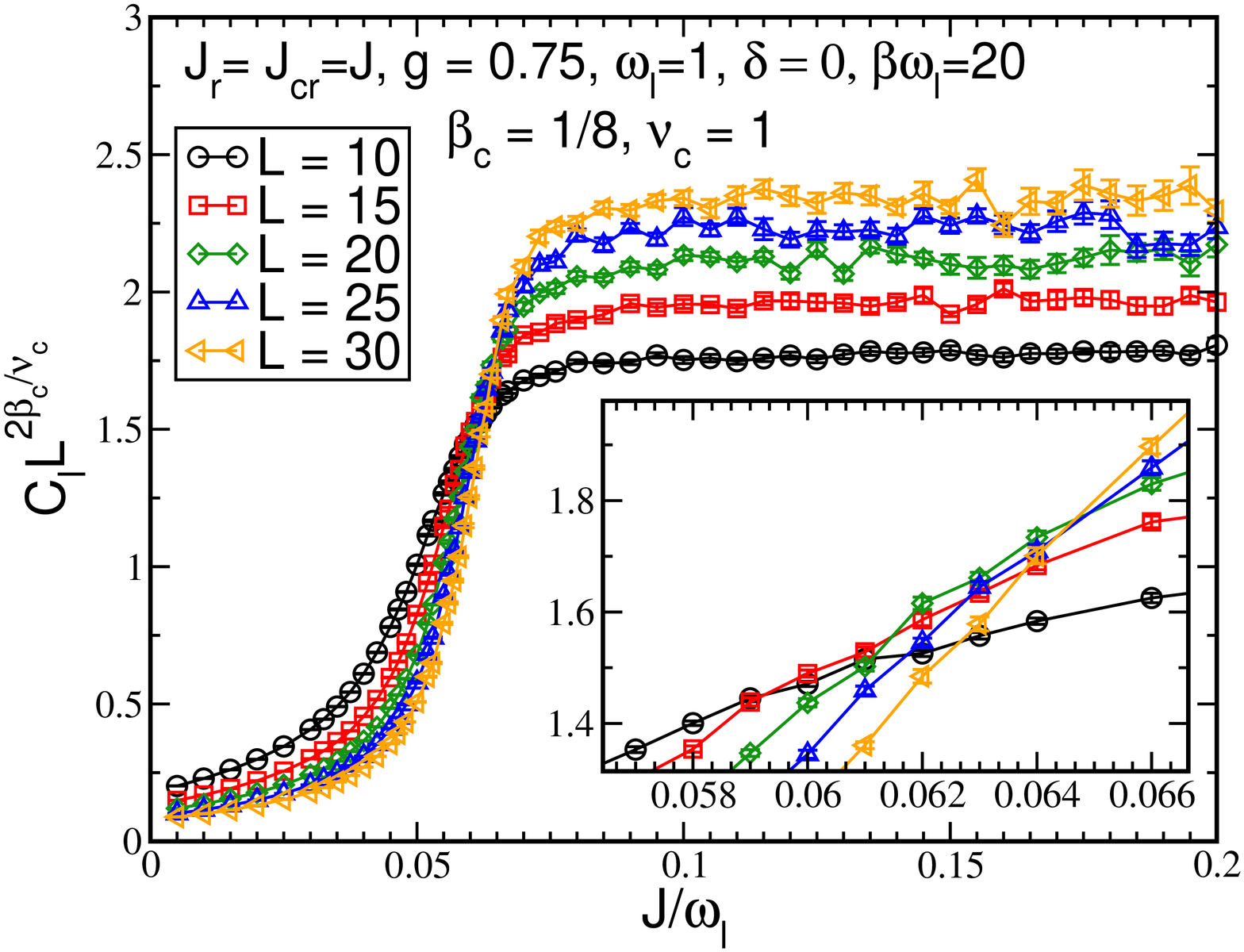}
        \caption{(Color online) Finite size scaling analysis at the
          coherent/incoherent transition. We used the critical
          exponents of the 2D Ising universality class and observe, as
          expected, a universal crossing of the curves. Inset: Zoom on
          the transition region. For this value of $g$, the transition
          is located around 
$J_c/\omega_l \simeq 0.062$.
\label{fss}}
\end{figure}

To determine more precisely the transition point between the
incoherent and coherent phases, we perform finite size scaling
analysis \cite{gould07}. This one-dimensional system has discrete
$Z_2$ symmetry and, consequently, its ground state quantum phase
transitions belong to the classical two-dimensional Ising model
universality class. The order parameter is $\langle a
\rangle=\sqrt{C_l}$ (or, equivalently, the magnetization of the spin
along the $x$-axis).  At the critical point, we expect the order
parameter to scale like $L^{-\beta_c/\nu_c}$ where $\beta_c = 1/8$ and
$\nu_c = 1$ are the critical exponents. Hence, $C_l
L^{2\beta_c/\nu_c}$ is independent of $L$ at the transition which
allows us to determine the critical point $J_c$ with good precision
(Fig. \ref{fss}). In addition, we confirmed (not shown
  here) the collapse of curves for different system sizes in the
  critical region by using the rescaled transition parameter
  $(J-J_c)L^{1/\nu_c}$. This confirms the two-dimensional Ising
  universality class of the transition.  Performing such analysis for
many different values of $g$ allows us to determine the transition
line between incoherent and coherent regions in the
$(J/\omega_l,g/\omega_l)$ plane.

 We also analysed this transition by doing the mean-field
  (MF) analysis previously performed by Schir\'o {\it et al.}
  \cite{schiro13}, which we brifely review here.  As the kinetic terms
  are the only ones that couple sites, we can replace them, for each
  site, with a coupling to an external homogeneous field $\psi =
  \langle a_i \rangle$ for all $i$. We are then left with a
  collection of identical on-site MF Hamiltonians
\begin{eqnarray}
  H^{MF}_i &=& \omega_s \sigma^+_i  \sigma^-_i + \omega_l a^\dagger_i a_i 
  + g \left(  \sigma^+_i  + \sigma^-_i\right) (a_i + a_i^\dagger) \nonumber \\
  && -J_r \left(\psi^\star a_i + \psi a^\dagger_i \right)  - J_{cr}
  \left(\psi a_i + \psi^\star a^\dagger_i \right). 
\end{eqnarray}
These on-site Hamiltonians can be numerically diagonalized, in the
standard way, by truncating a basis for a large enough number of
photons and a self consistent solution is obtained when the mean value
of $a_i$ calculated in the ground state is equal to the field $\psi$
inserted in the Hamiltonian. This also corresponds to a minimum value
of the energy with respect to $\psi$ as the method is variational.

To analyse the number of photons, we will now use an even
  simpler approximation, where we replace all photon operators $a_i$
  ($a^\dagger_i$) by one c-number $\psi$ ($\psi^\star$). This is a
  simple coherent state approach that should be valid in the coherent
  phase when the photon number is large enough to neglect
  fluctuations.  Our problem is then reduced to simple MF Hamiltonians
  $H^{SMF}_i$, with only two states, describing independent spins
  coupled to the same perfectly coherent field,
\begin{eqnarray}
  H^{SMF}_i &=& \omega_s \sigma^+_i  \sigma^-_i + \omega_l |\psi|^2
  + g \left(  \sigma^+_i  + \sigma^-_i\right) (\psi + \psi^\star) \nonumber \\
  && -2 J_r |\psi|^2 - J_{cr} \left(\psi^2 + (\psi^\star)^2\right).
\end{eqnarray}
Due to the $Z_2$ symmetry of the problem, the value of $\psi$ that
minimizes the energy can be chosen real and positive.  Solving this
two-state problem yields an expression for the transition line:
\begin{equation}
g = \frac{\sqrt{w_s(w_l -2 J_r -2J_{cr})}}{2 }\label{gcsimple},
\end{equation}
and the density of photons in the coherent region,
\begin{equation}
n_l = |\psi|^2 = \frac{g^2}{(\omega_l - 2J_r -2 J_{cr})^2} -
\frac{{\omega_s}^2}{16g^2}.\label{thelaw}
\end{equation}
This reduces to $n_l \simeq g^2 / (\omega_l - 2J_r -2J_{cr})^2 $ for
$\psi \gg 1$ where the approximation should be valid.  The number of
photons will then diverge for $w_l = 2(J_r + J_{cr})$ and the system
becomes unstable for $w_l$ smaller than this limit. Such instabilities
are observed in photonic systems, for example in \cite{koch09}, when
the chemical potential is large enough. Since $\omega_l$ plays the
role of a chemical potential in our system and as the number of
excitations is not conserved, it is reasonable to find such a
stability limit.

\begin{figure}[h]
    \includegraphics[width=8.8cm]{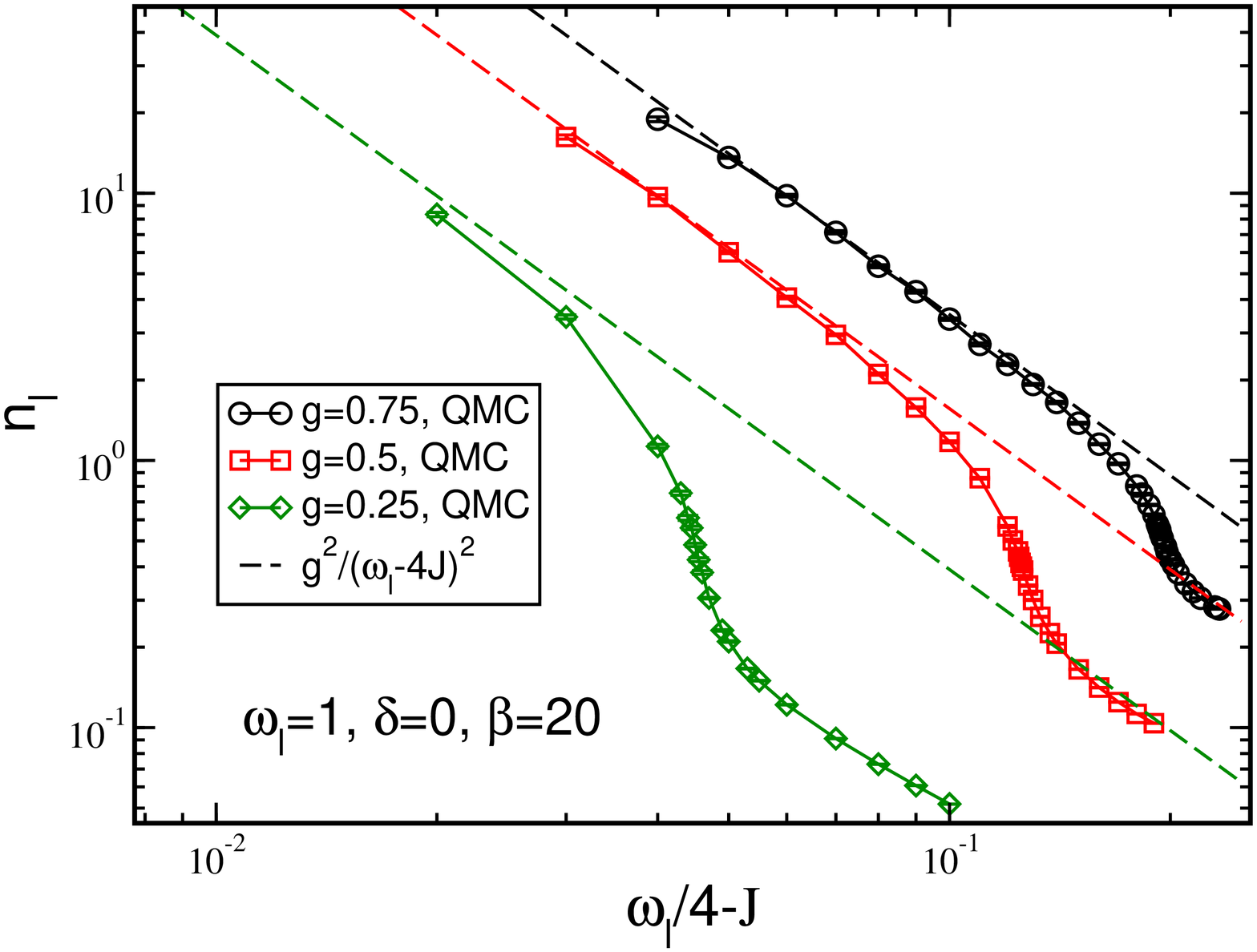}
        \caption{(Color online) Comparison between the expected divergence
of the density  of photons, $n_l$, predicted with a simple mean field
theory when $J \rightarrow \omega_l / 4$ and the QMC data. We find
very good agreement for intermediate densities. At larger densities,
the QMC becomes inefficient. Error bars are smaller than symbols.
\label{nphoton}}
\end{figure}

We used our QMC simulations to study the divergence of the number of
photons as the instability point is approached, Fig. \ref{nphoton}.
We observe that $n_l$ follows the simple MF result for the larger values of
the photon density, which is expected since our approximation is not
valid for small densities. It is very difficult numerically to get
closer to the divergence point because the QMC becomes inefficient
when there is a very large number of photons on a site (typically more
than ten or twenty photons per site).  However, there is always a
region where the MF behaviour is observed which confirms that the
system is unstable below a given value of $\omega_l$ ($4J$ in the case
of Fig. \ref{nphoton}).

\begin{figure}[h]
    \includegraphics[width=8.8cm]{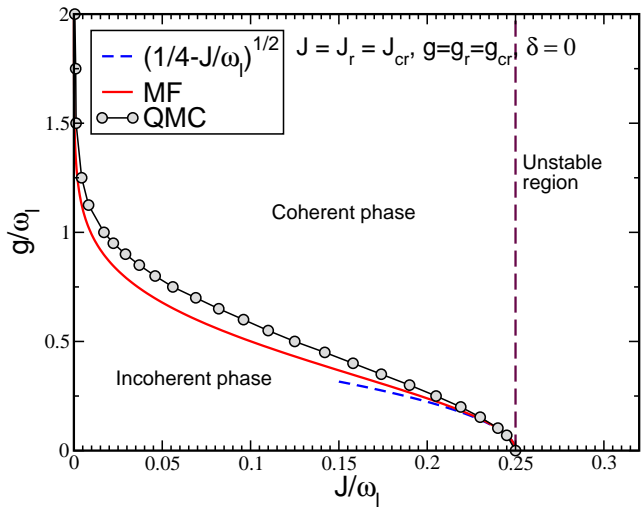}
    \caption{(Color online) The phase diagram of the RH model at zero
      temperature with $\delta = 0$ and $J_r = J_{cr} = J$.  There is
      an unstable region for $J/\omega_l > 1/4$. The rest of the
      diagram is separated between a coherent (``superradiant") region
      and incoherent (``insulating") one. The continuous red line
      reproduces the limits found in the MF study of Schiro {\it et
        al.} \cite{schiro13} and the blue dashed line corresponds to the simple expression
      Eq.(\ref{gcsimple}). Error bars are smaller than symbols.
 \label{diag}}
\end{figure}

As mentioned above, the finite size scaling analysis we performed for
different values of $g$, allows us to determine the phase diagram in
the $(J/\omega_l, g/\omega_l)$ plane which we show in Fig. \ref{diag}.

The MF results obtained
in \cite{schiro12,schiro13} provide a qualitatively correct
description of the transition line throughout the phase diagram.  On
the other hand, the simple ansatz Eq.(\ref{gcsimple}) only works well
for large $J$.

\begin{figure}[h]
    \includegraphics[width=8.8cm]{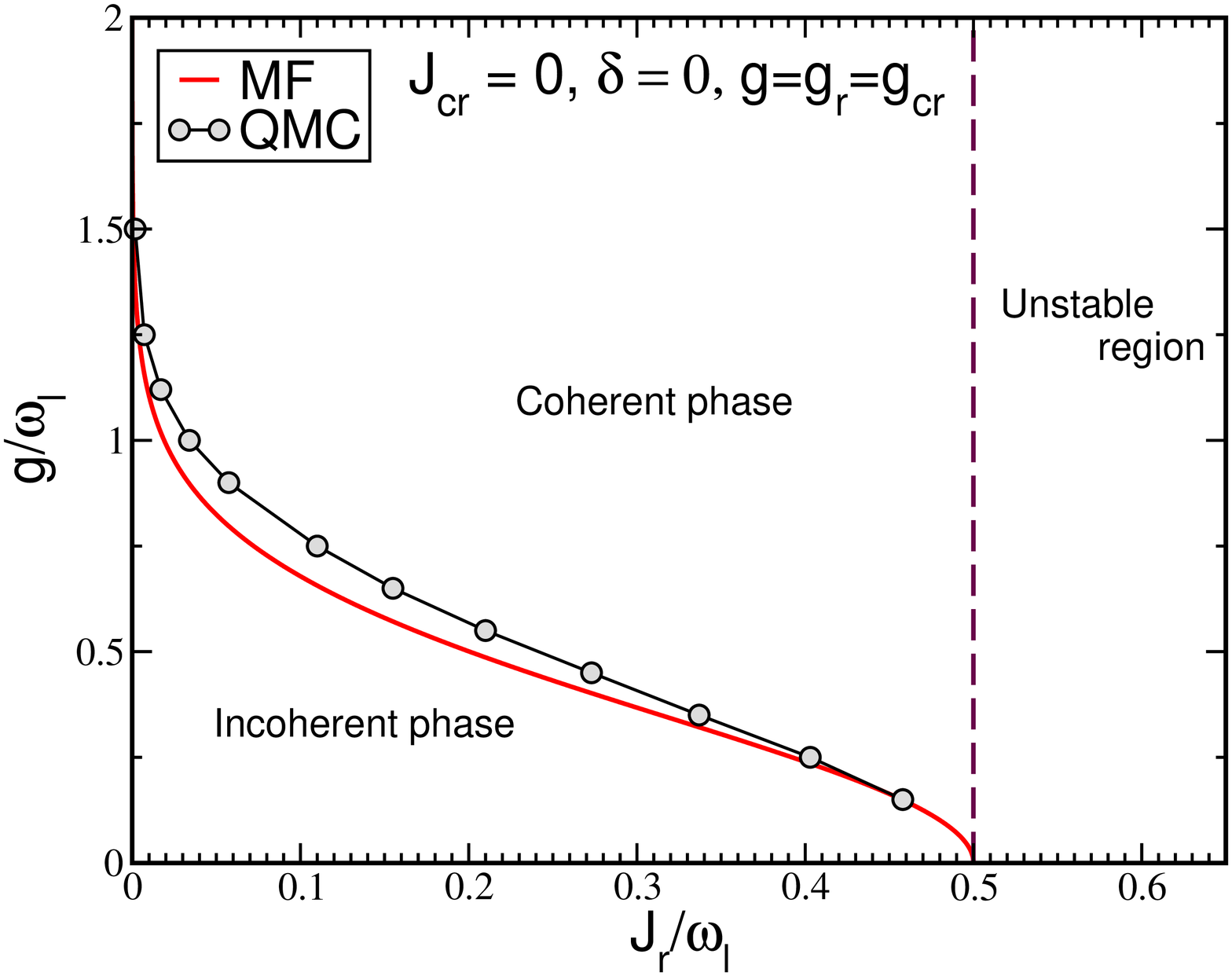}
        \caption{(Color online) The phase diagram of the RH model 
with $\delta=0$, $J_{cr} = 0$ and $g= g_r = g_{cr}$ is very similar to 
Fig. \ref{diag}. The continuous MF transition line was obtained
with the same technique as in \cite{schiro13}. Whenever we have
counter-rotating terms that are  large enough, whether in the hopping
term or in the exchange term between the spins and photons, we recover
qualitatively the same diagram with only two phases (incoherent and
coherent) and no Mott phases. Error bars are smaller than symbols.
\label{diagJr}}
\end{figure}

The absence of Mott insulating phases when the CR terms are taken into
account, predicted in \cite{schiro13}, is then also observed with our
QMC simulations. However the model used so far \cite{schiro13}
introduces unusual CR hopping, the $J_{cr}$ term in
Eq.(\ref{hoppingdetail}). In order to investigate the effect of this
term, we determined the phase diagram with ($J_r\ne 0, J_{cr} = 0$)
while maintaining the CR term in the on-site exchange ($g_r = g_{cr} =
g$).  We find a qualitatively similar phase diagram shown in
Fig. \ref{diagJr}.  This demonstrates that, when the counter-rotating
term is large enough, be it in the hopping or the exchange terms, the
Mott insulating phases are suppressed. This can be understood by
noting that the CR terms couple states with different numbers of
excitations and, consequently, forbid the photon blockade
\cite{birnbaum05} which is needed for the Mott phases to form
\cite{koch09}.

\section{Effects of the counter-rotating terms}

In the preceding section, we focused on the case with zero detuning,
and found that the CR terms played a major role in the physics of our
system and could not be neglected. This was expected as the values of
$g$ we used are large ($g \simeq \omega_l$). We will now study cases
where the detuning, $\delta$, is non zero or where $g_{cr}$ is smaller
than $g_r$ to see if CR terms play an important part in these cases
too.  To simplify our study we will keep only the on-site CR term and
set $J_{cr}=0$.  We will use the MF approximation that was used in
\cite{schiro13}, as it provides qualitatively correct results, as well
as QMC.

\begin{figure}[h]
\includegraphics[width=8.8cm]{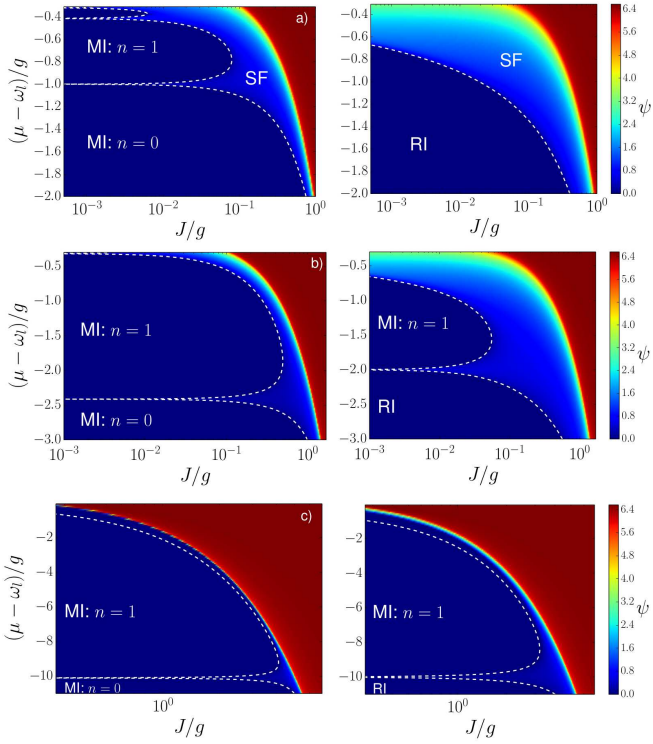}
\caption{(Color online) The MF phase diagrams of the JCH (left column) and RH (right column) models
for different $\delta$ at zero temperature. The color scale represents
the real positive value of the MF order parameter $\psi$.
(a) At $\delta =0$, the phase diagrams are very
different due to the degeneracy of the spin and photon energies. The
JCH model shows several Mott insulator (MI) phases with different
densities $n$, whereas the RH model only shows its incoherent/coherent
transition. 
(b) As $\delta/g = -2$ is decreased from zero, the 
MI phases with $n>1$ shrink in the 
JCH model whereas a small $n=1$ MI phase appear in the 
RH model. (c) At large $\delta/g = -10$, the two phase diagrams
are almost equivalent with a single visible $n=1$ MI phase.
\label{diagd0}}
\end{figure}

We start by comparing the MF phase diagrams of the JCH and RH models
at zero detuning, Fig. \ref{diagd0} (a), in the $(J/g, (\mu-\omega_l)/g)$
plane, where $\mu$ is the chemical potential. For the JCH model
$g_{cr} = 0$, $g = g_r$ and we vary $\mu-\omega_l$ which acts as the
effective chemical potential for the photons in the grand canonical
ensemble. For the RH model $g_r = g_{cr} = g$ and we tune $\omega_l$,
setting $\mu=0$.  The notion of canonical or grand canonical ensemble
is not really pertinent for the RH model as the number of excitations
is not a conserved quantity.  As mentioned before, the models have
very different phase diagrams.  If we introduce positive detuning,
$\delta > 0$, that is if $\omega_s > \omega_l$, we favor photon over
spin excitations and this leads to similar differences between the two
models as we encountered for $\delta=0$.

However, if we introduce negative detuning $\delta < 0$, the physics
of the two models becomes equivalent, as shown in Fig. \ref{diagd0} (b) and (c).
This is because $\delta <0$ favors atom excitations over photon creations
and, then, reduces the possible fluctuations of the total number of excitations.
In both models, we observe a region with no
excitations, a Mott insulator phase for $n=1$. However, we did not
observe Mott phase at $n=2$ in the RH model.  As $|\delta|$ becomes larger (but always
negative), the two phase diagrams become more similar as the MI phases
with $n>1$ in the JCH model disappear and as the shape of the $n=1$
MI phase becomes more similar between the two models (Fig. \ref{diagd0} (c)).

\begin{figure}[h]
    \includegraphics[width=8.8cm]{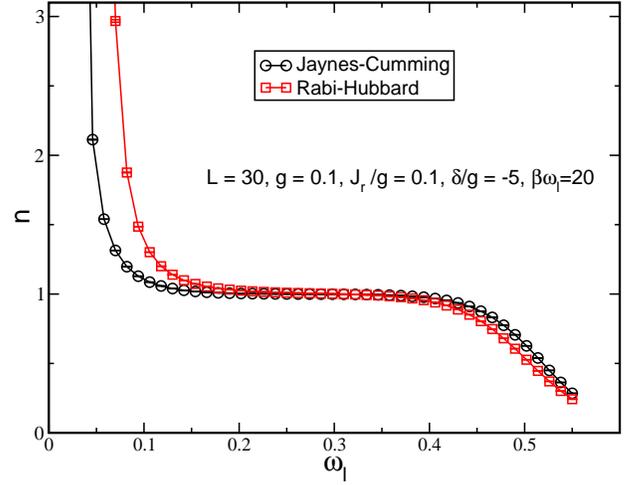}
        \caption{(Color online) Density $n$ of excitations versus $\omega_l$ for
a fixed $g$ and $J_r$ for the RH and JCH models using QMC. In both
cases, we observe the Mott like plateau for a density $n=n_s+n_l=1$ of
excitations with similar width. In both cases we do not observe the
$n=2$ plateau. Error bars are smaller than symbols.
\label{nvsw}}
\end{figure}

To confirm
this MF prediction, we show some QMC cuts in the phase diagram for
negative detuning in Fig. \ref{nvsw}. We observe the similarity of the
behavior in this case with the $n=1$ Mott plateau visible in the JCH
and the RH model, with similar width and the absence of a visible
$n=2$ plateau.  We then conclude that, for large values of $g$, the
counter-rotating terms can be neglected only if there is a negative
detuning that is large enough to overcome the coupling between states
with different densities of excitations. 

\begin{figure}[h]
    \includegraphics[width=8.8cm]{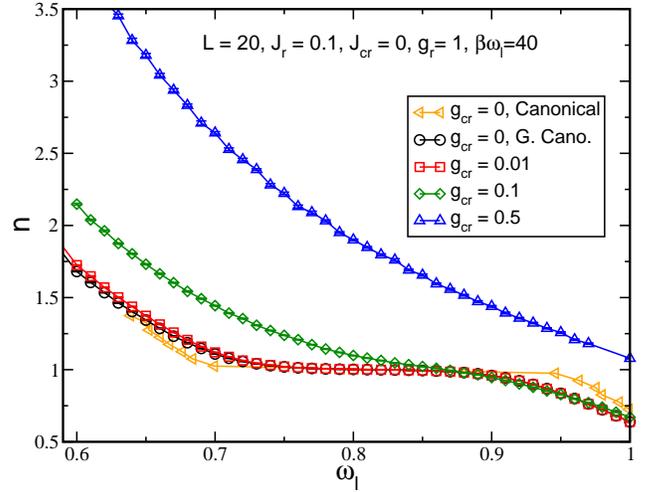}
        \caption{(Color online) Number of excitations $n$ as a function
of $\omega_l$ for different values of $g_{cr}$ and fixed $J_r$ and
$g_r$, for the Jaynes-Cummings and Rabi-Hubbard models. We see that
the two models give similar results for $g_{cr}/g_r=0.01$ but that the
Mott plateau is destroyed even for $g_{cr}/g_r=0.1$. Error bars are smaller than symbols.
\label{diffgcr}}
\end{figure}

We now consider directly the effects of the counter-rotating term by
tuning the value of $g_{cr}$ keeping the other parameters fixed. We
will use the zero detuned case (see Fig. \ref{diffgcr}). As expected,
we find that for small values, $g_{cr}/g_{r} = 0.01$, the RH model
gives results that are similar to the JCH model (treated in the
canonical or grand canonical ensemble).  However for $g_{cr}/g_{r}=0.1$, the CR terms
already have a strong effect and modify the physics by destroying the
$n=1$ Mott phase. We see, then, that the physics is very sensitive to
the addition of small counter-rotating terms since, even with $g_{cr}$
an order of magnitude smaller than $g_r$, we obtain a profound
modification of the phase diagram.

\newpage

\section{Conclusion}

We studied the Rabi-Hubbard model and compared it with the
Jaynes-Cummings model to see how the presence of counter-rotating
terms, that involve fluctuations of the number of photons, changes
the physics of said models.

For zero detuning, we observed that for large values of the
counter-rotating terms, be they in the hopping or on-site terms, the
physics of the two models is completely different as the Mott phases
present in the Jaynes-Cummings are destroyed in the RH model. We also
observed that, due the discrete $Z_2$ symmetry of the RH model, there
is a transition between an incoherent phase and phase with true long
range coherence similar to the Dicke transition,
and we identified an unstable region in parameter
space. These results are in good qualitative agreement with the MF
results previously obtained by Schiro {\it et al.} \cite{schiro12,schiro13}.

Exploring further, we identified two ways in which the effects of the
counter-rotating terms can be reduced: by introducing a negative
detuning or by directly making the counter-rotating parameter $g_{cr}$
smaller. In both cases, the physics of the RH model approaches that of
the JCH model.

By using arbitrarily large values of the counter-rotating terms, we
depart from most experimental systems that have very small $g$
parameters
Even with rather small values of $g$, the counter rotating terms
 can have a big impact on the phase
diagrams, as is also observed in ultra strong coupling experiments
\cite{niemczyk10,forndiaz16}.

 Another important effect in experiments is the presence
  of dissipative terms: losses from the cavity and gains to compensate
  those. To obtain a strong coupling regime between the two-state
  system and the photons, these terms are generally small compared to
  $g$. However the system is driven out of equilibrium because of such
  effects.

  Whereas we expect a Bose-Einstein condensation of photons at
  equilibrium, when the system is out of equilibrium, we expect phase
  coherence to appear through a lasing effect. The question of the
  difference between a laser and a Bose-Einstein condensate for
  photons is the subject of current debate \cite{schmitt15}, with the
  suggestion that there is a crossover between these two forms of
  coherent states.  The QMC technique we use does not allow to study
  non Hermitian Hamiltonians but, including gain/loss terms that are
  Hermitian conjugates, we can address this question.

\acknowledgments

\noindent
The authors would like to thank B. Gr\'emeaud and C. Miniatura for interesting discussions.

\medskip

\noindent
All the authors contributed equally to the paper.

\end{document}